\documentclass[amsmath,amssymb, aps, physrev,10pt]{revtex4-2}

\usepackage{graphicx}
\usepackage{dcolumn}
\usepackage{bm}
\usepackage{amsmath}
\usepackage{color}
\usepackage{MnSymbol}
\usepackage{subfigure}
\usepackage{ragged2e}
\usepackage{float}
\usepackage{silence}
\usepackage{hyperref}

\def\xvec{\mathbf{x}}
\def\islinkedto{\prec \hspace{-1.5mm}*\hspace{0.5mm}}

\begin{document}

\title{\textbf{Numerical Evaluation of the Causal Set Propagator in 2D Anti-de Sitter Spacetime }} 
\author{Arsim Kastrati}
\email{Contact author: arsim.kastrati@uni-wuerzburg.de}
\author{Haye Hinrichsen}%
\affiliation{Faculty for Physics and Astronomy, Julius Maximilians University Würzburg, Am Hubland, 97074 Würzburg, Germany}

\date{\today}

\begin{abstract}
We numerically investigate the application of the path-sum-based causal set scalar propagator construction to (1+1)-dimensional Anti-de Sitter (AdS) spacetime. Building upon a generalization of Johnston's path sum approach, we simulate Poisson-sprinkled causal sets in AdS$_{1+1}$ and numerically evaluate the retarded scalar propagator, comparing it to the known continuum result. Our results confirm that even in curved spacetimes with constant negative curvature, the discrete causal set path sum reproduces the continuum propagator without modification of the flat-spacetime jump amplitudes, thereby providing further numerical support for former analytical results and the applicability of the path sum formalism to curved Lorentzian manifolds.
\end{abstract}

\maketitle

\section{Introduction}
%
Formulating a consistent theory of quantum gravity remains one of the foremost open problems in theoretical physics. The challenge arises from the conceptual tension between quantum field theory (QFT), which assumes a fixed spacetime background, and general relativity, in which spacetime itself is dynamic and subject to quantum fluctuations. Bridging this gap requires a fundamental rethinking of spacetime, particularly at the Planck scale, where the smooth manifold structure of general relativity is expected to break down due to quantum effects. This breakdown is often signaled by ultraviolet (UV) divergences in quantum field theories of gravity, suggesting the need for a microscopic or discrete structure underlying the continuum approximation of spacetime~\cite{tHooft1974,Carlip2001}.

Several approaches have been proposed to address this challenge, including Loop Quantum Gravity (LQG)~\cite{Rovelli2004}, Causal Dynamical Triangulations (CDT)~\cite{Ambjorn2006}, String theory~\cite{Green1987}, and holographic dualities such as the AdS/CFT correspondence~\cite{Maldacena1998,Witten1998}. LQG and CDT share a fundamentally discrete view of spacetime and introduce a minimal length scale directly in their formulation. String theory, though formulated on a smooth manifold, effectively incorporates a minimal length through the finite size of strings. 

Following from String theory, the AdS/CFT correspondence offers a duality between a gravitational theory in $(d+1)$-dimensional Anti-de Sitter (AdS) spacetime and a conformal field theory (CFT) living on its $d$-dimensional boundary. In particular, the AdS/CFT framework does not \textit{a priori} assume a discrete structure or a fundamental length scale, raising questions about its compatibility with discrete or emergent spacetime models. This motivates the study of \textit{discrete holography}, which seeks to reconcile the AdS/CFT correspondence with theories positing a fundamental discreteness of spacetime. Although efforts using isometric lattices and hyperbolic tessellations have demonstrated promising techniques for discretizing hyperbolic spaces~\cite{basteiro,magnus1974noneuclidean,schrauth2024hypertilinghighperformance}, these constructions often lack a dynamical notion of time or causal structure. Moreover, such lattices require a high degree of symmetry such as constant negative curvature, and thus cannot capture arbitrary configurations of the gravitational field. 

To avoid these difficulties, an alternative approach may come from Causal Set Theory (CST), which postulates that the underlying structure of spacetime is a locally finite partially ordered set, where the order reflects the causal relations between spacetime events~\cite{Sorkin2005,Bombelli1987,Rideout1999}. Causal sets provide a framework in which Lorentzian geometry emerges from a fundamentally discrete causal structure, bypassing many of the difficulties associated with UV divergences in quantum field theory. In CST, a continuum spacetime can be approximated by ``sprinkling'' points via a Poisson process over the manifold and then extracting their causal relations. The causal set itself retains no geometric information beyond the order relation, and yet in the continuum limit it can encode key geometric quantities such as spacetime volume and curvature~\cite{Dowker2005,Henson2006,Glaser2014}.

Although much of the research in CST has focused on flat Minkowski spacetime (see, e.g.~\cite{Johnston2008,Dowker2013,Shuman_2024}), there have also been numerous promising studies in curved embedding spaces~\cite{benincasa2010scalar,Belenchia_2016,X_2017,nomaan2021aspects,ashmead2023estimating,homvsak2024boltzmannian}, including de Sitter and Anti-de Sitter~\cite{Glaser2013,Buck2013,X_2017,Surya_2019,Philpott2009}. These play a central role in cosmology and holography, respectively.  In particular, AdS$_{1+1}$ spacetime serves as a minimal yet non-trivial testing ground to explore how causal set dynamics and field propagation behave in negatively curved backgrounds.

In this work, we contribute to this line of inquiry by numerically studying the retarded scalar field propagator on a causal set approximation of (1+1)-dimensional AdS spacetime. By comparing the discrete propagator with its continuum counterpart, we aim to provide additional numerical evidence for the applicability of CST path sum constructions in curved backgrounds, supporting analytical results found by Nomaan X et. al. \cite{X_2017}. Our results indicate that causal set methods can be meaningfully extended to curved spacetimes, showing that the geometry is fully captured by the order relations and the density of causal events.

We begin by establishing the necessary theoretical background, reviewing the fundamentals of Causal Set Theory and the geometry of AdS$_{1+1}$, including its conformal structure, isometries, and geodesics. We recall how the continuum retarded propagator is derived from the Klein-Gordon equation using geodesic distance and eigenfunction expansion techniques. On the discrete side, we review Johnston and Shuman’s path sum approach, which also provides a robust framework for constructing discrete propagators in curved spacetimes. We then perform numerical simulations of the discrete propagator in causal sets sprinkled in AdS$_{1+1}$. By comparing the discrete result to the known continuum propagator, we demonstrate strong agreement and highlight the effectiveness of the CST approach in capturing scalar field dynamics in curved Lorentzian geometries.

\section{Theoretical framework}
To construct a scalar-field propagator on a discrete causal set embedded in a Lorentzian manifold such as AdS spacetime, we first summarize the essential frameworks and mathematical tools. In this Section we begin with a review of the basic definitions of CST as a discrete spacetime model, where causal relations define the fundamental structure.  Next, we outline the geometry of AdS$_{1+1}$, focusing on its conformal structure, isometries, and geodesics. For determining the continuum propagator, we then turn to the ordinary Klein-Gordon equation, expressing its solutions explicitly in terms of geodesic distances in Minkowski and AdS spacetime. To complete the discussion about continuum scalar field theory, we go over the method of eigenfunction expansion, which is a general method to find the propagator in eigenspace. Finally, for the discrete case, we recap Johnston's "Hop-and-Stop" model \cite{johnston2010quantumfieldscausalsets} and its extension by Shuman~\cite{Shuman_2024} for constructing discrete propagators in a more general form.

\subsection{Brief Introduction to Causal Set Theory}
%
A \textit{causal set} $\mathcal{C}$ is a locally finite partially ordered set $(C, \prec)$. The set $C$ represents a set of events in spacetime, which is equipped with a binary relation $\prec$ such that for all events $x, y, z \in C$ the following axioms are satisfied:
\begin{enumerate}
    \item[(i)] \textbf{Causality (Partial Order Condition):} There exists a binary relation $\prec$ such that for any two elements $x, y \in C$, if $x \prec y$, then $x$ is in the causal past of $y$.
    \item[(ii)] \textbf{Transitivity:} If $x \prec y$ and $y \prec z$, then $x \prec z$.
    \item[(iii)] \textbf{Antisymmetry:} If $x \prec y$ and $y \prec x$, then $x = y$.
    \item[(iv)] \textbf{Local Finiteness (Discrete Spacetime):} For any pair of elements $x, y \in C$, the set of causally connected elements between them, called the \textit{causal interval} [$x,y$]$=\{z \in\mathbb C \mid x \prec z \prec y\}$, is finite.
\end{enumerate}
In a certain sense, a causal set is both Lorentzian [(ii),(iii)] and discrete [(iv)]. Each of the above properties plays a fundamental role in defining the structure of spacetime \cite{johnston2010quantumfieldscausalsets}.

To analyze how large causal sets approximate the continuum structure of a Lorentzian manifold, it is essential to define the concept of an \textit{embedding}. Given a causal set $(C, \prec)$, an embedding in a Lorentzian manifold $(M, g)$ is a mapping $ f: C \to M $ that preserves causal order, which means $ x \prec y $ if and only if $ f(x) \prec f(y) $. Furthermore, an embedding is said to be \textit{faithful} if the images of the elements of the causal set under $ f $ are uniformly distributed in $ M $ according to its volume measure $\sqrt{-g}$. An important conjecture in causal set theory, known as the \textit{Hauptvermutung}, asserts that if a causal set admits a faithful embedding into two different spacetimes $(M, g)$ and $(M', g')$, then these spacetimes must be approximately isometric \cite{Surya_2019}.

A faithful embedding into a given Lorentzian manifold can be created by a method known as \textit{sprinkling}. Sprinkling refers to randomly placing points in the Lorentzian manifold according to a Poisson distribution. This means that the probability $P(n,V)$ of finding $ n $ points within a space-time volume $ V $ is given by
\begin{align}
P(n,V) = \frac{(\rho V)^n}{n!} e^{-\rho V},
\end{align}
where $ \rho $ represents the sprinkling density \cite{Roy_2013}. The Poisson distribution ensures statistical invariance under isometries such as Lorentz boosts \cite{DOWKER_2004}.

In a causal set, a continuous trajectory is replaced by a discrete sequence of causally connected events. These sequences can be classified as follows:
\begin{itemize}
\item
A \textbf{chain} in a causal set $\mathcal{C} = (C, \prec)$ is defined as a totally ordered subset of elements $\{x_0, x_1, ..., x_n\} \subseteq C$ such that $x_0 \prec x_1 \prec ... \prec x_n$. Chains generalize the notion of a causal curve in spacetime and can be thought of as discrete analogs of worldlines.
\item
A \textbf{link} between two elements $x,y\in C$, denoted as $x \islinkedto y$, is a causal relation with no other causally connected elements between, i.e., there does not exist another element $z \in C$ such that $x \prec z \prec y$. Two events connected by a link may be thought of as causally connected nearest neighbors.
\item
A \textbf{path} between two elements $x, y \in C$ is a sequence $\{x_0 = x, x_1, ..., x_n = y\}$ of linked elements such that each consecutive pair satisfies $x_i \islinkedto x_{i+1}$. The maximal number of elements of all paths naturally provides a way to define a distance between $x$ and $y$, analogous to the length of a timelike geodesic line for which the integrated proper time is maximal. \cite{sorkin2003causalsetsdiscretegravity}.
\end{itemize}
A finite causal set can be conveniently represented using an \textit{adjacency matrix}. Labeling the events by  $ x_1, \dots, x_p $, two $ p \times p $ different adjacency matrices are defined, namely, the \textit{causal matrix} $A$ and the \textit{link matrix} $L$
with the components
\begin{equation}
A_{ij} :=
\begin{cases}
1, & \text{if } x_i \prec x_j \\[-1mm]
0, & \text{otherwise}
\end{cases}
\,, \qquad
L_{ij} :=
\begin{cases}
1, & \text{if } x_i \islinkedto  x_j \\[-1mm]
0, & \text{otherwise}
\end{cases}\,.
\end{equation}
Both matrices have zeros along their main diagonal. Furthermore, it is always possible to choose a labeling such that both matrices are strictly upper triangular. When such a labeling is used, it is referred to as a \textit{natural labeling} \cite{PhysRevD.101.065013}. The powers of these matrices have useful properties \cite{johnston_2025_99pg0-gkb40}:
\begin{align}
(A^n)_{ij} &= \text{number of chains of length } n \text{ from } x_i \text{ to } x_j,\\[2mm]
(L^n)_{ij} &= \text{number of paths of length } n \text{ from } x_i \text{ to } x_j.
\end{align}
These are the most important definitions of CST, for further details, see, e.g.,~\cite{dribus2013axiomscausalsettheory}.

\subsection{2D Anti de-Sitter Spacetime}
%
In this work, we intend to demonstrate that the CST is ideally suited to approximate arbitrarily curved manifolds in a discrete representation. As a simple example, we will consider an Anti-de Sitter spacetime in $(1+1)$-dimensions (AdS$_{1+1}$), which plays an important role in the study of the AdS/CFT correspondence~\cite{ammon2015gauge}. In the following, we briefly summarize its essential features.

AdS$_{1+1}$ is a Riemannian manifold with Lorentzian signature and constant negative Gaussian curvature $K=-\frac{1}{L^2}$, where $L>0$ is the AdS curvature radius. This manifold can be represented in various coordinate systems. In the following, we use conformal coordinates
\begin{equation}
\xvec=(x^0,x^1)=(t,x ) \qquad \text{with} \qquad
t\in\mathbb R,\;\;
x \in(-\tfrac\pi 2,+\tfrac\pi 2)\,.
\end{equation}
In these coordinates, the metric tensor of AdS$_{1+1}$ is given by
\begin{equation}
g_{\mu\nu}(\xvec)=\frac{L^2}{\cos^2 x }\eta_{\mu \nu} \,,
\label{metricads}
\end{equation}
where $\eta_{\mu\nu}$ denotes the Minkowski metric. The corresponding volume element reads
\begin{equation}
\label{volumelement}
\sqrt{-g} = \frac{L^2}{\cos^2 x }\,.
\end{equation}
Note that in a conformal metric, the causal relation is the same as in Minkowski space, namely,
\begin{equation}
\label{CausalOrder}
x \prec y \qquad \Leftrightarrow \qquad x^0 < y^0 \; \wedge \; |x^1-y^1|< |x^0-y^0|   \,.
\end{equation}
The metric in Eq. \eqref{metricads} and its volume element are preserved under isometries generated by the Killing vector fields
\begin{equation}
\begin{split}
X_T &\;=\; \partial_t\\
X_S &\;=\; -\sin t\sin x  \;\partial_t + \cos t\cos  x  \;\partial_x  \\
X_B &\;=\;  \cos t\sin x  \;\partial_t + \cos x \sin t \;\partial_x\,,
\end{split}
\end{equation}
corresponding to translations in time (T), the analog of translations in space (S), and boost transformations (B). Explicitly, these isometries are given by
\begin{equation}
\label{isometries}
\begin{split}
T_\tau(t,x)   &=  (t+\tau,x )\\
S_\alpha(t,x) &=
\Bigl(\text{arccot}\bigl( \cosh\alpha\cot t + \sinh\alpha\tfrac{\sin x}{\sin t} \bigr),\,
      \arctan\bigl( \cosh\alpha\tan x + \sinh\alpha\tfrac{\cos t}{\cos x} \bigr) \Bigr)\\
B_\eta(t,x)   &=
\Bigl(\arctan\bigl( \cosh\eta\tan t+ \sinh\eta\tfrac{\sin x}{\cos t}  \bigr),\,
      \arctan\bigl( \cosh\eta\tan x + \sinh\eta\tfrac{\sin t}{\cos x} \bigr)\Bigr)\,,
\end{split}
\end{equation}
where $\tau,\alpha,\eta$ are real parameters.

In AdS$_{1+1}$ with the metric~\eqref{metricads} the geodesic equations are
\begin{equation}
\ddot t + 2\tan( x) \,\dot t \,\dot  x = 0
\; , \qquad
\ddot  x + \tan( x) \bigl(\dot t^2+\dot x^2\bigr) = 0\,.
\end{equation}
Solving these equations leads to an explicit expression for the geodesic line connecting the two points $\xvec_1=(t_1, x_1)$ and $\xvec_2=(t_2, x_2)$, that is,
\begin{equation}
\begin{split}
 x(t) &= \arcsin\bigl( A \sin(t-t_0) \bigr) \qquad \ \text{for timelike geodesics }(|A|<1)\\[2mm]
 x(t) &= \pm(t-t_0) \hspace{26mm} \text{for lightlike geodesics } (|A|=1)\\
t( x) &= t_0+\arcsin\bigl( \frac{1}{A} \sin( x) \bigr) \qquad \text{for spacelike geodesics }(|A|>1)\\
\end{split}
\end{equation}
with the constants
\begin{align}
A^2 \;&=\; \frac{\sin^2  x_1+\sin^2 x_2-2\cos(t_2-t_1)\sin x_1\sin x_2}{\sin(t_2-t_1)}\\
t_0 \;&=\;
\arctan\Bigl( \frac{\sin t_1 \sin x_2-\sin t_2\sin  x_1}{\cos t_1\sin x_2-\cos t_2 \sin x_1} \Bigr)\,.
\end{align}
The length of a geodesic line in AdS$_{1+1}$, known as \textit{geodesic distance}, is known to be given by
\begin{equation}
\label{geodesicdistance}
d(\xvec_1,\xvec_2)\;=\;
\begin{cases}
L \arccos\Bigl( \frac{\cos(t_1-t_2)-\sin x_1\sin x_2}{\cos x_1\cos x_2} \Bigr)
& \quad \text{for timelike (causal) connections} \\
L \, \text{arccosh}\Bigl( \frac{\cos(t_1-t_2)-\sin x_1\sin x_2}{\cos x_1\cos x_2}\Bigr)
& \quad \text{for spacelike connections} \\
0
& \quad \text{for lightlike connections}\\[1mm]
\text{undefined}
& \quad \text{if $\frac{\cos(t_1-t_2)-\sin x_1\sin x_2}{\cos x_1\cos x_2}<-1$}
\end{cases}
\end{equation}
As expected, the geodesic distance turns out to be invariant under the isometries in Eq. \eqref{isometries}.

\subsection{Continuum Scalar Field Theory}
\label{sft}
%
For a CST to be physically meaningful, it must be possible to define appropriately discretized field theories that can approximate the known field-theoretical results in the continuum. Here, we will restrict ourselves to a simple scalar field theory, for which we briefly summarize the known results in the continuum.

\subsubsection{Retarded Propagators in Minkowski and Anti de-Sitter Spacetime}
Let us first consider a scalar field $\phi(\xvec)$ with mass $m\geq 0$ in a Minkowski space in ($d$+1) dimensions. It evolves according to the Klein-Gordon equation $\bigl(\Box -m^2\bigr)\phi=0$, where $\Box=\partial_\mu\partial^\mu = \delta-\partial_t^2$ is the d'Alembert operator using the ``mostly plus'' convention. For the Klein-Gordon equation, the retarded two-point propagator coincides with the Green function $G(\xvec,\xvec')$, which obeys the differential equation
\begin{equation}
\bigl(\Box -m^2\bigr) G(\xvec,\xvec') \;=\; -\delta^{d+1} (\xvec-\xvec')\,.
\end{equation}
Invariance under translations, rotations, and Lorentz boosts implies that $G$ depends only on the temporal relativistic distance $\tau=\sqrt{-x_\mu x^\mu}=\sqrt{t^2-\vec x^2}$, turning the above differential equation into
\begin{equation}
\label{MinkowskiDE}
\bigl(\frac{1}{\tau^d} \partial_\tau \tau^d \partial_\tau + m^2\bigr)G(\tau) \;=\;
G''(\tau) + \frac{d}{\tau}G'(\tau) + m^2G(\tau) \;=\; 
\frac{\delta(\tau)}{S_{d-1}\tau^d}\,,
\end{equation}
where $S_{d-1}=2\pi^{d/2}/\Gamma(d/2)$ is the surface area of a sphere in $d$ dimensions. In  dimensions (1+1), the solution of this differential equation is known to be given by~\cite{johnston2010quantumfieldscausalsets}
\begin{equation}
G(\tau) \;=\; \theta(t) \theta(\tau^2)\frac12 J_0(m \tau)\,,    
\end{equation}
where $J_0$ denotes the Bessel function of the first kind and $\theta(\tau)$ is the Heaviside step function. 

Turning to AdS$_{d+1}$, the propagator is determined by the partial differential equation
\begin{equation}
(\Box_g -m^2)G(\xvec',\xvec)\;=\;-\frac{\delta^{d+1}(\xvec-\xvec')}{\sqrt{-g}}\,.
\end{equation}
where $\Box_g = \frac{1}{\sqrt{-g}}\partial_\mu \sqrt{-g}\,g^{\mu \nu }\partial_\nu$ is now the Laplace-Beltrami operator that generalizes the d'Alembert operator to curved spaces. Again, the isometries~\eqref{isometries} tell us that $G$ can only depend on the causal (that is, temporal) geodesic distance $\tau=d(\xvec',\xvec)$, leading to the differential equation
\begin{equation}
\label{GeneralDE}
\Bigl(\frac{1}{\sin^d(\tau/L)}\partial_\tau \sin^d(\tau/L) \partial_\tau +m^2\Bigr)G(\tau) \;=\;
G''(\tau)+\frac{d}{L \tan(\tau/L)}G'(\tau)+m^2G(\tau)
\;=\; 
\frac{\delta(\tau)}{S_{d-1}\,L^d\sin^d(\tau/L)} \,.
\end{equation}
Note that in the limit $\tau\to 0$ this differential equation reduces to Eq.~\eqref{MinkowskiDE}. This can be used to fix the normalization of the solution. In $(1+1)$-dimensions $(d=1)$ the solution of Eq. \eqref{GeneralDE} reads
\begin{equation}
G(\tau)\;=\;  \frac12 \theta(t) \theta(\tau^2) P_\ell(\cos\frac\tau L),
\label{contpropads}
\end{equation}
where $P_\ell(z)$ is the Legendre function and $\ell=\frac12\bigl( \sqrt{1+4L^2m^2}\,-1\bigr)$. In the massless case $m=0$ this solution reduces to
\begin{equation}
G(\tau)\;=\;  \frac12 \theta(t) \theta(\tau^2)
\end{equation}
which is constant inside the future light cone.
\subsubsection{Expansion in the eigenmodes of the Laplace-Beltrami operator}
\label{ritus}
When calculating propagators in Minkowski space, the underlying translation invariance is usually exploited by Fourier-transforming the differential equation. This allows the problem to be represented in the basis of the eigenfunctions of $\Box$ as a plane-wave expansion. However, as we will see below, the eigenmode decomposition can be useful when ordinary translation invariance is absent and even when the eigenmodes are not explicitly known. This situation occurs, for example, in Sturm-Liouville problems but also in AdS spaces to be examined below~\cite{ritus1978,correa,habermann2021asymptoticerroreigenfunctionexpansion}.

To outline this approach, let us consider a linear differential operator $ \mathcal{L} $, which governs the equation of motion $\mathcal{L} \Phi = 0$ of a scalar field $\Phi(\xvec)$. The corresponding propagator $K(x,y)$ satisfies
\begin{equation}
    \mathcal{L} K(x, y) = \pm \delta(x - y).
    \label{greensfunc}
\end{equation}
Solving the eigenvalue equation 
\begin{equation}
    \mathcal{L}   \Phi_\lambda (x) = \lambda \Phi_\lambda (x),
    \label{eigenvalueeq}
\end{equation}
yields a set of eigenvalues $\lambda$ with the corresponding eigenfunctions $ \Phi_\lambda (x) $. If these eigenfunctions form a complete basis, they satisfy the orthogonality relation
\begin{equation}
    \int dx \, \Phi_\lambda (x) \Phi^*_{\lambda'} (x) = 
    \begin{cases}
\delta(\lambda - \lambda') & \text{for continuous } \lambda  \\
\delta_{\lambda, \lambda'} & \text{for discrete } \lambda
\end{cases}
\label{ortho}
\end{equation}
and the completeness relation
\begin{equation}
    \sumint d\lambda \, \Phi_\lambda (x) \Phi^*_{\lambda'} (y) \;=\; \delta(x - y)\,,
    \label{complete}
\end{equation}
where the symbol $\sumint$ stands for a sum over discrete or an integral over continuous eigenvalues, respectively. With Eqs.~\eqref{greensfunc}, \eqref{eigenvalueeq}, and \eqref{complete} one can easily show that the propagator can be expanded in this basis as
\begin{equation}
    K(x, y) = \sumint d\lambda \, \Phi_\lambda (x) \tilde{K} (\lambda) \Phi^*_\lambda (y)\,,
\end{equation}
where
\begin{equation}
    \tilde{K} (\lambda) = \pm \frac{1}{\lambda}.
\end{equation}
As we will see in the following, this approach will be useful even if we do not know the eigenfunctions $\Phi_\lambda$ explicitly. 

\subsection{Construction of Discrete Propagators}
%
In quantum theory, propagators describe the amplitude for a system to transition between two states or events. For a relativistic particle, the Feynman path integral formalism expresses this amplitude as a sum over all possible trajectories between two spacetime points, each weighted by $ e^{iS[q]} $, where $ S[q] $ is the action along the path $ q $~\cite{RevModPhys.20.367}. Although not all propagators can be derived in this way, those for free scalar particles coincide with the Green’s functions of the Klein-Gordon equation~\cite{PhysRevD.48.748}.

In the causal set framework, where spacetime is fundamentally discrete and causal relations are encoded combinatorially, the concept of a path sum is adapted by assigning amplitudes to discrete trajectories. Johnston~\cite{Johnston_2008,johnston2010quantumfieldscausalsets} proposed a model in Minkowski space in which each jump in a trajectory contributes a \emph{hop amplitude} $ a $, and each intermediate event contributes a \emph{stop amplitude} $ b $. Summing over such paths yields a propagator of the form
\begin{equation}
K(x, y) = \sum_n a^n b^{n-1} P_n(x, y),
\end{equation}
where $ P_n(x, y) $ is the average number of paths of length $ n $ between the events $ x $ and $ y $ in the causal set. This expression can be reformulated using convolution integrals and solved via Fourier transforms to recover continuum results for scalar field propagators on average.

\begin{figure}[t!]
    \centering
    \includegraphics[width=0.25\linewidth]{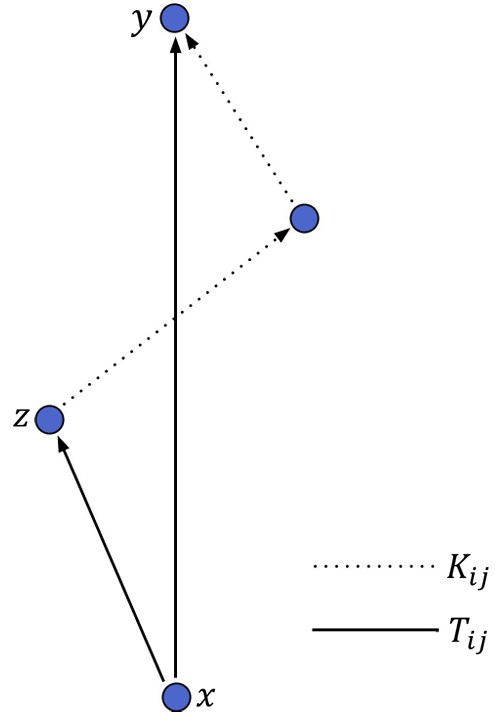}
    \caption{\justifying Illustration of the path sum approach. Every trajectory from  $x$ to $y$ is either a direct transition to $y$, or it involves an intermediate event $z$, followed by a path from $z$ to  $y$. This is captured in the composition rule in Eq. \eqref{cstpropagator}.
}
    \label{fig:pathsum}
\end{figure}

\subsubsection{Discrete Propagators and Jump Amplitudes }
%
The approach described above is not unique. Other models, such as in \cite{scargle2009photondispersioncausalsets}, feature non-constant hop amplitudes without stop amplitudes. Alternatively, \cite{sorkin2007doeslocalityfailintermediate,Belenchia_2015,Aslanbeigi_2014} define propagators as Green’s functions of a causal set d’Alembertian operator, completely bypassing the summation over paths.

The diversity of constructions that yield correct continuum behavior on average suggests a broader family of viable path sum formulations. Shuman’s recent work \cite{Shuman_2024} yields results analogous to those of Johnston, although expressed in a more general framework. In particular, the relation between jump amplitudes and scalar field propagators in causal sets is formulated so as to recover the continuum behavior after averaging over Poisson sprinklings into Minkowski space. In the following, we summarize this approach and note that it remains applicable in broader settings, including curved spacetimes.

As we already know, in a finite causal set, each event can be labeled by an index. Define a matrix $ T $, where $ T_{xy} $ is the probability amplitude for a direct jump from event $ x $ to event $ y $. The total amplitude for all paths from $ x $ to $ y $ is denoted $ \sigma_{xy} $, which includes all sequences of such jumps, see Fig.~\ref{fig:pathsum}. The propagator $ K_{xy} $ is proportional to $ \sigma_{xy} $
\begin{equation}
\label{ProportionalityConstant}
K_{xy} = \alpha \, \sigma_{xy},
\end{equation}
Since all trajectories can be built from sequences of jumps, we get the recursion relation
\begin{equation}
\sigma_{xy} = T_{xy} + \sum_z T_{xz} \sigma_{zy}.
\label{compositionlaw}
\end{equation}
This leads to the matrix equation
\begin{equation}
K = \alpha T + TK \quad \Rightarrow \quad K = \alpha T(I - T)^{-1},
\label{cstpropagator}
\end{equation}
where the matrix inverse $ (I - T)^{-1} $ is guaranteed  to exist \cite{Johnston_2008,johnston2010quantumfieldscausalsets}. This provides a well defined formula for computing the propagator in terms of the jump amplitude matrix $ T $.

\subsubsection{Sprinkling Averages}
To compare causal set propagators with continuum results, one averages over all Poisson sprinklings on the manifold, defining averaged functions $ T(x, y) $ and $ K(x, y) $ as mean values of the matrix elements $ T_{xy} $ and $ K_{xy} $ for all sprinklings containing the events $ x $ and $ y $.  These averages should follow a composition law similar to Eq.~\eqref{compositionlaw}. However, in the continuum limit, the sum over discrete events $z$ becomes an integral over spacetime. Since each summand contributes an average ``discreteness volume'' $ V_0 = 1/\rho $, we can approximate the sum as an integral
\begin{equation}
\sum_z T_{xz} K_{zy} \;=\; \rho \sum_z T_{xz} K_{zy} V_0 \; \approx \; \rho \int dz \, T(x,z) K(z,y).
\end{equation}
Combining this with the averaged version of $ K_{xy} $, we obtain the continuum relation
\begin{equation}
K(x,y) \;=\; \alpha T(x,y) + \rho \int dz \, T(x,z) K(z,y).
\label{convo}
\end{equation}
In the case of translation invariance (e.g. in Minkowski space), where $T(x, y) = T(y - x)$ and $K(x, y) = K(y - x)$, this integral becomes a convolution
\begin{equation}
K(y - x) \;=\; \alpha T(y - x) + \rho \int dz \, T(z - x) K(y - z).
\end{equation}
This convolution equation can be conveniently solved in momentum space. Applying a Fourier transform, we get
\begin{equation}
\tilde{T}(p) = \frac{\tilde{K}(p)}{\alpha + \rho \tilde{K}(p)}.
\label{jumpampp}
\end{equation}
This shows how the average propagator relates to the average jump amplitude in momentum space. By Fourier transforming back to position space, one obtains the form of the jump amplitudes that, when summed over all paths and averaged over sprinklings, reproduce the continuum propagator. 

In curved spacetimes like AdS$_{1+1}$, however, translational invariance is replaced by non-commuting isometries and hence a standard Fourier decomposition cannot be used. In the following, we show that the method of path sums still holds in general, making use of the fact presented in Sec. \ref{ritus}, which assumes that every function can be expanded in terms of a complete and orthogonal set of eigenfunctions of the respective operator.

Let us assume that there is a complete set of eigenfunctions $\Phi_\lambda(x)$ and that $T(x,y)$ and $K(x,y)$ can be expanded
\begin{align}
T(x, y) =& \sumint d\lambda \, \Phi_\lambda (x) \tilde{T} (\lambda) \Phi^*_\lambda (y), 
\\ 
K(x, y) =& \sumint d\lambda \, \Phi_\lambda (x) \tilde{K} (\lambda) \Phi^*_\lambda (y).
\end{align}
Plugging these expansions into Eq.~\eqref{convo}, we obtain after some rearrangements
\begin{equation}
\begin{split}
\sumint d\lambda \, \Phi_\lambda (x) \tilde{K} (\lambda) \Phi^*_\lambda (y) &= \, \alpha \sumint d\lambda \, \Phi_\lambda (x) \tilde{T} (\lambda) \Phi^*_\lambda (y) \\
 & + \, \rho \sumint d\lambda d\lambda' \, \Phi_\lambda (x) \tilde{T} (\lambda)\tilde{K} (\lambda') \Phi^*_{\lambda'} (y)\int dz \Phi_{\lambda'} (z)\Phi^*_\lambda (z). 
\end{split}
\end{equation}
Now we can use the orthogonality relation in Eq. \eqref{ortho} to evaluate the integral over $z$, leading to
\begin{equation}
\begin{split}
\sumint d\lambda \, \Phi_\lambda (x) \tilde{K} (\lambda) \Phi^*_\lambda (y) = \, \alpha &\sumint d\lambda \, \Phi_\lambda (x) \tilde{T} (\lambda) \Phi^*_\lambda (y)  \\
+ \, \rho  &\sumint d\lambda \, \Phi_\lambda (x) \tilde{T} (\lambda)\tilde{K} (\lambda) \Phi^*_{\lambda} (y).
\end{split}
\end{equation}
Finally, since the eigenfunctions are linearly independent, the above expression must also hold for the coefficients only, giving
\begin{equation}
\tilde{K}(\lambda) = \alpha \tilde{T}(\lambda) + \rho \tilde{T}(\lambda) \tilde{K}(\lambda) \quad \Rightarrow \quad \tilde{T}(\lambda) = \frac{\tilde{K}(\lambda)}{\alpha + \rho \tilde{K}(\lambda)}\,.
\end{equation}
This is almost the same expression for the jump amplitude as in Eq. \eqref{jumpampp}, except that it lives in a more general eigenfunction space.
\subsubsection{Determining the Jump Amplitude for a given Propagator}
\label{solvjumpamp}
From Sec. \ref{ritus} we know that scalar field propagators (retarded, advanced, and Feynman) in the eigenfunction space typically have the form
\begin{equation}
\tilde{K}(\lambda) = \frac{1}{f(\lambda) + m^2}.
\end{equation}
Inserting this into the previous expression for $ \tilde{T}(\lambda) $, we find
\begin{equation}
\tilde{T}(\lambda) = \frac{1}{\alpha} \cdot \frac{1}{f(\lambda) + m^2 + \rho/\alpha}.
\label{fourierjump}
\end{equation}
This resembles a propagator with an effective mass shift
\begin{equation}
m^2 \mapsto m^2 + \frac{\rho}{\alpha}.
\end{equation}
If we introduce a scaling factor
\begin{equation}
\beta = \sqrt{1 + \frac{\rho}{m^2 \alpha}},
\end{equation}
to write this shift compactly, the final relation for the average jump amplitude, after applying an inverse Fourier transformation to Eq. \eqref{fourierjump}, becomes
\begin{equation}
T(x,y) = \frac{1}{\alpha} K(x,y)|_{m \, \to \, \pm\beta m}
\label{jumpamp}
\end{equation}
The result shows that the implementation of a jump amplitude leads basically to a renormalization of the mass parameter. Consequently, a massive propagator can be obtained from a massless one by implementing a jump amplitude. Moreover, this formulation enables the construction of entire families of jump amplitudes which, on average, reproduce known continuum propagators through a discrete path sum on causal sets. As we will show in the next Section, this approach can be used to find a discrete propagator also in AdS$_{1+1}$ spacetime.

\section{Numerical Evaluation and Results}
%
In this Chapter, we test whether the discrete causal set propagator matches the known continuum retarded propagator in AdS$_{1+1}$ spacetime. We begin by outlining suitable methods for sprinkling points in this case, taking into account the constant curvature and the causal structure. We then derive the appropriate jump amplitude by following the procedure presented in Sec. \ref{solvjumpamp}. With these tools, we numerically calculate the discrete propagator for arbitrary curvature radius $L$ and compare it to the continuum result. 

\subsubsection{Sprinkling into AdS$_{1+1}$ }
\label{sprinklingads}
To generate a sprinkling, one randomly distributes $N$ points with uncorrelated positions on the given manifold at constant density $\rho$, which means that the probability density of finding a point is proportional to the volume element~$\sqrt{-g}$. The number of points is chosen such that $N=\lfloor\rho V\rfloor$, where $V$ is the total volume or area in which the sprinkling is generated.

\begin{figure}[t]
\includegraphics[width=130mm]{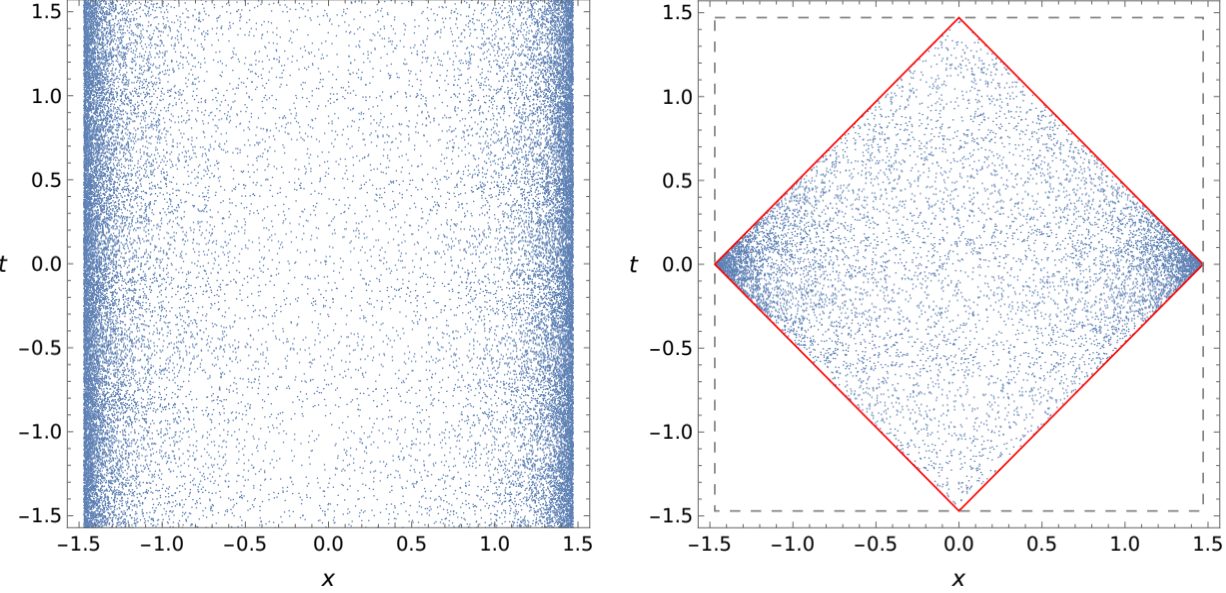}
\caption{\justifying Sprinkled points in AdS$_{1+1}$ with density $\rho=1000$ and cutoff $\epsilon=0.1$. Left: Simple method which generates points in the entire strip $(-\tfrac\pi 2,\tfrac\pi 2)\times (-\tfrac\pi 2+\epsilon,\tfrac\pi 2-\epsilon)$. Right: Improved method which generates points exclusively inside a rhombus (red line). The cutoff is indicated as a dashed line. }
\label{fig:sprinkling}
\end{figure}

Attempting to create sprinklings on AdS$_{1+1}$ throughout the entire range $(t,x)\in(-\tfrac\pi 2,\tfrac\pi 2)\times(-\tfrac\pi 2,\tfrac\pi 2)$, one encounters the problem that the total volume is infinite since $\sqrt{-g}=\tfrac{L^2}{\cos^2 x}$ diverges at the conformal boundary $x \to\pm\tfrac\pi 2$. Therefore, one is forced to implement a cutoff and to restrict the spatial coordinate to the range $x\in(-\tfrac\pi 2+\epsilon,\tfrac\pi 2-\epsilon)$ with $0<\epsilon \ll 1$, so that the total volume
\begin{equation}
V\;=\; L^2\int_{-\pi/2}^{\pi/2}\text{d} t \int_{-\pi/2+\epsilon}^{\pi/2-\epsilon}\frac{1}{\cos^2 x}\text{d} x
\;=\; \frac{2\pi L^2}{\tan\epsilon}
\end{equation}
is finite. To randomly sprinkle a point, two uniformly distributed random numbers $z_1,z_2 \in [-1,1]$ are generated, and one defines its coordinates by
\begin{equation}
t\;:=\;\pi z_1
\,,\qquad
x\;:=\;\arctan\bigl(\frac{z_2}{\tan\epsilon} \bigr)\,.
\end{equation}
This process is repeated until $N=\rho V$ points have been generated. A typical sprinkling is shown in the left panel of Fig.~\ref{fig:sprinkling}. Note that the points are distributed according to Poisson statistics, which automatically ensures invariance under the isometries of the underlying manifold.

When numerically calculating a two-point function, it is important to consider only those pairs of points for which the implementation of the cutoff has no effect, since otherwise uncontrollable systematic errors may occur. Therefore, the numerical evaluation must be limited to pairs of points such that the intersection of the forward light cone of the starting point and the backward light cone of the target point does not touch the cutoff. This can be achieved, for example, by limiting the calculation to the rhombus-shaped area and discarding all points outside it. However, this means that most of the points, namely all those outside the rhombus, are not needed. To avoid creating and destroying a large number of points for nothing, we propose an equivalent, improved method which works as follows. We generate two uniformly distributed random numbers $z_1,z_2\in(-1,1)$, set $w:=(\sin\epsilon)^{|z_1|}$, and define the coordinates by
\begin{equation}
t:=\text{sign}(z_1) \bigl(\arcsin(w)-\epsilon \bigr)
\,,\qquad
x:=\arctan\Bigl( \frac{\sqrt{1-w^2}}{w}z_2 \Bigr)\,.
\end{equation}
Again, this process is repeated until $N=\rho V_R$ points have been generated, where $V_R \;=\; -4 L^2\ln(\sin\epsilon)$ is the volume of the rhombus. An example of the resulting sprinkling is shown in the right panel of Fig.~\ref{fig:sprinkling}.

The improved sprinkling method is significantly faster because it generates fewer points near the conformal boundary. For a given density $\rho$, this results in a reduction in execution time by a factor of the order $\tfrac{V_R}{V} = -\tfrac 2 \pi \epsilon \ln \epsilon$, which is particularly advantageous when $\epsilon$ is small. For example, for $\epsilon=10^{-2}$, the improved sprinkling method is about 34 times faster.

\subsubsection{Jump Amplitudes in AdS$_{1+1}$}
%
As shown in~\cite{Shuman_2024} in more detail, the proportionality constant $\alpha$ introduced in Eq. \eqref{ProportionalityConstant} can be found by dimensional analysis. It depends on the mass $m$, the density $\rho$, and a free parameter $\gamma$:
\begin{equation}
\alpha =\gamma \frac{\rho}{m^2} \quad \Rightarrow \quad \beta = \sqrt{1 + \frac{1}{\gamma}}\;.
\end{equation}
\begin{figure}[t!]
    \centering
    \subfigure[]{\includegraphics[width=0.45\textwidth]{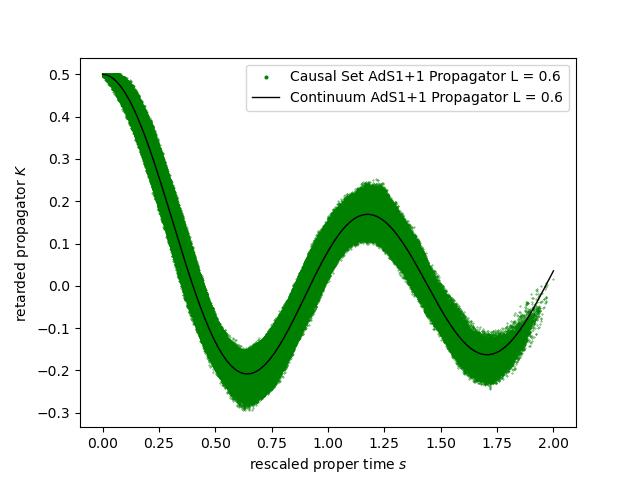}}
    \hspace{0.0001\textwidth}
    \subfigure[]{\includegraphics[width=0.45\textwidth]{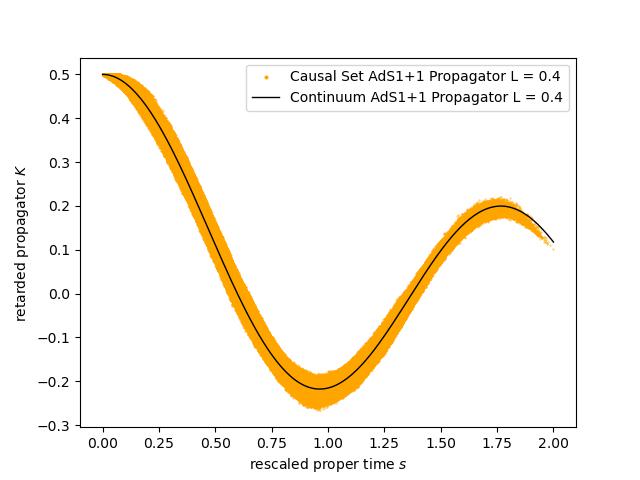}}
    \hspace{0.0001\textwidth}
    \subfigure[]{\includegraphics[width=0.45\textwidth]{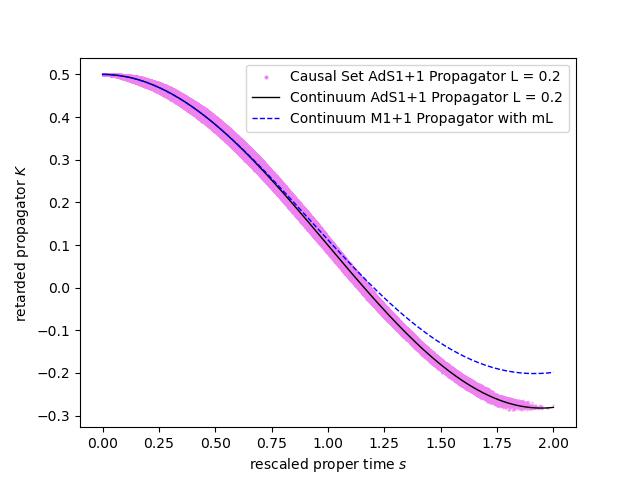}}
    \hspace{0.0001\textwidth}
    \subfigure[]{\includegraphics[width=0.45\textwidth]{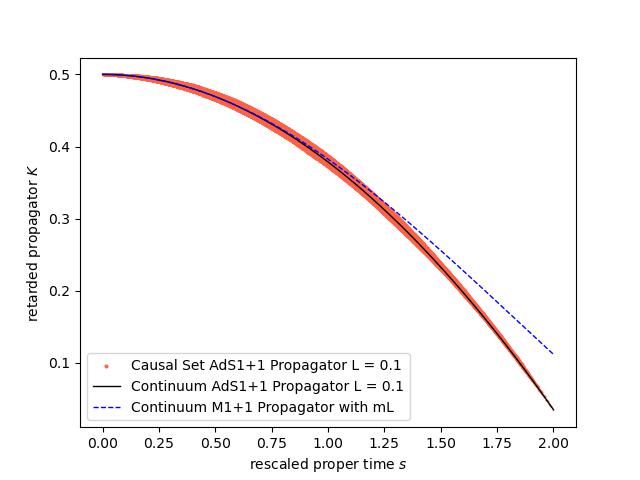}}
    \caption{\justifying Numerical evaluation of the retarded causal set propagator in Eq.~\eqref{cstprop} for $m=10$, obtained from a single sprinkling of $n=18000$ points. Panels (a)–(d) show results for $\mathrm{AdS}_{1+1}$ with different curvature radii $L$. In each case, the causal set propagators are presented as functions of the rescaled manifold proper time and compared with the corresponding continuum retarded propagators. In panels (c) and (d), the flat-space continuum propagator with effective mass $m_{\text{eff}} = mL$ is additionally included (dotted curve) for comparison.}
    \label{fig:propresultsfigs}
\end{figure}
From Eq.~(\ref{jumpamp}) we know that the jump amplitude $T(x,y)$ should match on average the scalar field propagator in AdS$_{1+1}$ shown in Eq. \eqref{contpropads}, giving
\begin{equation}
T(x,y) \;=\; \frac{m^2}{\gamma\rho} \frac{1}{2} \Theta(t) \Theta(s^2) P_{\ell}(\cos s) \quad \text{with} \quad \ell = \frac{1}{2}(-1+\sqrt{1+4 \beta^2 m^2 L^2}),
\end{equation}
where $s=\frac{\tau}{L}$ is just the rescaled proper time. Since $\gamma$ can be chosen freely, we can set $\gamma = -1$ so that $\beta = 0$, thereby relating the jump amplitude to the massless causal propagator
\begin{equation}
     T(x,y) \;=\; -\frac{m^2}{2\rho} \Theta(t) \Theta(s^2),
     \label{ampjumpamp}
\end{equation}
which is constant inside the future light cone. Note that this result coincides with the one in the Minkowski case in $(1+1)$-dimensions \cite{Shuman_2024,johnston2010quantumfieldscausalsets}.
This is not surprising, as the AdS$_{1+1}$ spacetime is topologically trivial and can be described using conformally flat coordinates. Thus, the retarded propagator for a massless field has to be the same as in a flat spacetime \cite{X_2017}.

\subsubsection{Numerical Results }
In order to demonstrate the feasibility of the method numerically, we begin by selecting events in the causal set using Poisson sprinkling, as outlined in Sec.~\ref{sprinklingads}, with $N = 18000$ points distributed within a causal diamond in a $(1+1)$-dimensional Anti-de Sitter spacetime. The causal order of the resulting set does not depend on the conformal factor of the metric in Eq. \eqref{metricads} and can therefore be determined by using the Minkowski metric (cf. Eq.~\eqref{CausalOrder}). With that, the causal matrix $A_{xy}$ can be calculated. This causal matrix will then be used to construct a jump amplitude matrix $T(x, y)$. Recall that $T(x, y)$ was defined as the average of $T_{xy}$ over all sprinklings. Consequently, for the causal set propagator to agree with the continuum propagator on average, the ensemble average of $T_{xy}$ must reproduce the continuum solution Eq. \eqref{ampjumpamp}. By setting $\alpha = -\frac{m^2}{\rho}$, the jump amplitude matrix is defined as
\begin{equation}
    T_{xy} = -\frac{m^2}{2 \rho} A_{xy}.
    \label{finaljumpamp}
\end{equation}
Putting this into Eq.~\eqref{cstpropagator}, we obtain the propagator on the causal set
\begin{equation}
K_{xy} = \frac{1}{2} A_{xy} \Bigl(I +  \frac{m^2}{2 \rho} A_{xy} \Bigr)^{-1}\,.
\label{cstprop}
\end{equation}
Note that this expression coincides with the one obtained in the $(1+1)$-dimensional Minkowski spacetime \cite{johnston2010quantumfieldscausalsets} up to the fact that the causal matrix is different. This observation confirms that all information about the curvature of the embedded spacetime is encoded solely in the causal order and the local density of the events.  

To demonstrate that the retarded causal set propagator correctly approximates the continuum behavior, we numerically calculate $K_{xy}$ for a single sprinkling and a fixed mass $m=10$ for various curvature radii $L$ and compare the results with the corresponding continuum solutions. The results are shown in Figs. \ref{fig:propresultsfigs} and  \ref{fig:boringpropsfigs}.  Every dot in the plots represents a pair of causally connected sprinkled points, plotting the value of the propagator against their (rescaled) proper time difference on the respective manifold.

As can be seen, the numerical analysis of retarded causal set propagators in $\mathrm{AdS}_{1+1}$ spacetime reveals excellent agreement with the corresponding continuum solutions over a wide range of curvature radii $L$. The propagator curves obtained from causal set sprinklings closely track their continuum counterparts, supporting the claim that discrete causal dynamics can faithfully reproduce continuum field propagation in curved spacetimes.

One key feature of our analysis is that the number of sprinkled elements was kept fixed while the AdS curvature scale $L$ was varied. Since the total volume of the spacetime region is proportional to $L^2$, the sprinkling density $\rho$ scales as $\rho \propto 1/L^2$. As a consequence, smaller values of $L$ correspond to higher densities and thus smaller discrete fluctuations. This trend is clearly visible in the data: for smaller $L$, the discrete propagators adhere more tightly to the continuum curve, while at larger $L$, increased sparsity leads to more pronounced fluctuations. This behavior confirms the expected relationship between the sprinkling density and the fidelity of the causal set propagator.

To provide a benchmark, we also included results from the $(1+1)$-dimensional Minkowski case in Fig. \ref{fig:boringpropsfigs} (a). This serves both as a consistency check and a baseline for comparison. As anticipated, the causal set propagator in AdS$_{1+1}$ with $L = 1$ (Fig. \ref{fig:boringpropsfigs} (b)) closely matches the flat space result, reflecting the fact that curvature effects are minimal at this scale and further confirming the accuracy of the construction in the weakly curved regime.

To investigate the sensitivity of the causal set propagator to curvature and to test whether it captures the correct geometry, we additionally compare it to a flat space propagator with an \emph{effective mass} term given by $m^2_{\text{eff}} = m^2 L^2$, as shown in Fig. \ref{fig:propresultsfigs} (c)-(d). This serves as an analytic approximation to the AdS propagator for small proper times $s$, where the effects of curvature are approximately encapsulated in an effective mass shift. The comparison reveals that this propagator approximates the AdS continuum solution well for small $s$, validating its use as a local approximation.

Our numerical results support previous analytic findings by Nomaan X, Dowker and Surya \cite{X_2017}, who showed that in two-dimensional AdS spacetime the \emph{hop-and-stop constants} of the causal set scalar propagator remain identical to their flat space values within a \emph{Riemannian Normal Neighborhood (RNN)}. In such regions, the continuum is effectively flat and curvature corrections only enter at higher orders. Our numerical results confirm that, even without approximations, the full AdS$_{1+1}$ causal set propagators reproduce the expected continuum behavior with the same constant amplitudes as in flat space. This provides direct numerical support for the analytic finding that causal order, together with the appropriate sprinkling density, suffices to encode the curvature effects of AdS$_{1+1}$ spacetime.

\section{Conclusion}
%
In this work, we have provided additional numerical evidence for the applicability of CST path sum constructions in curved backgrounds, focusing specifically on the $(1+1)$-dimensional Anti-de Sitter spacetime. By implementing a suitable Poisson sprinkling procedure and using a path sum approach, we numerically calculated a discrete retarded scalar field propagator and compared it with the known continuum counterpart derived from the Klein-Gordon equation. Our results exhibit remarkable agreement of the discrete and continuum propagators across a wide range of AdS curvature scales, without requiring any modification to the flat-space jump amplitudes, affirming the ability of causal sets to faithfully reproduce continuum dynamics, even in the presence of curvature. This agreement not only validates the robustness of the CST framework in curved settings, but also illustrates how curvature effects are faithfully encoded in the causal order and local sprinkling density alone, confirming the previous analytical result found in~\cite{X_2017}. This also confirms Johnston's assumption that the model and the corresponding amplitudes in flat space may not have to be changed to correctly reproduce the curvature dependence of the continuum Green’s functions \cite{johnston2010quantumfieldscausalsets}.

Our results presented here open several promising avenues for extension in the direction of quantum field theory and possibly toward a deeper understanding of holography in discrete spacetimes. To extend causal networks with classical fields to quantum-mechanical degrees of freedom, several promising directions have been taken. The Sorkin–Johnston vacuum provides a key step in this direction. It constructs a unique, covariantly defined vacuum state based on the retarded and advanced Green's functions of the d’Alembertian on causal sets. Recent developments have extended this construction even to massive fields and de-Sitter spacetime \cite{Mathur_2019,Surya_2018}. The formulation of interacting quantum field theories on causal sets remains an open frontier. There is emerging work on defining $\phi^4$ and other interacting theories using perturbative expansions and Green's functions on causal sets \cite{albertini2021phi4}. A different strategy is the use of algebraic quantum field theory (AQFT) techniques on causal sets, which focus on the algebra of observables avoiding to introduce a fixed background metric \cite{Dable_Heath_2020, wreo29866,Fewster_2021,Fewster_2024}.

Another promising frontier is the implementation of discrete holography with CST. The construction of boundary theories using techniques such as aperiodic spin chain duals from hyperbolic tilings has already provided discrete counterparts to the AdS/CFT correspondence \cite{basteiro}. These developments suggest that holography may not be exclusive to smooth manifolds but could emerge from fundamentally discrete structures. It is also worth mentioning that new links have been found between AdS/CFT and AQFT \cite{Witten_2022}. In this context, our demonstration that causal set propagators accurately reflect the AdS curvature without altering jump amplitudes reinforces the feasibility of defining consistent bulk-boundary maps in a fully discrete setting.

Nevertheless, the presented approach is not without limitations. Firstly, the method still relies on sprinkling, a process that randomly embeds elements into a continuum manifold according to a Poisson distribution. While this preserves Lorentz invariance and enables consistent discretization, it also implicitly assumes the existence of the underlying continuum geometry. This reliance means that the method still lacks a truly dynamical or background-independent notion of spacetime. Various approaches have been proposed to address this issue, such as classical sequential growth models and transitive percolation techniques \cite{Rideout1999}, which attempt to generate causal sets intrinsically and dynamically. However, significant challenges remain, particularly in reproducing continuum-like curved manifolds from such intrinsically discrete dynamics. Understanding the emergence of geometric properties from purely combinatorial growth remains an active and difficult area of research.

Secondly, although the propagator-based approach is appealing due to its avoidance of traditional differential operators, a complete and predictive theory of quantum gravity based on causal sets will inevitably require analogs of standard calculus tools. In particular, substantial progress has been made in this direction. Discrete d'Alembertian operators have been developed for causal sets \cite{sorkin2007doeslocalityfailintermediate}, enabling the definition of dynamics for scalar fields, including massive ones, and incorporating curvature corrections, marking a major step toward defining differential operators in a discrete setting \cite{Benincasa_2010,Belenchia_2016}. Furthermore, causal set versions of action and variational principles have been proposed \cite{Benincasa_2011,Roy_2013,Bombelli_2021}, opening the way to eventually formulate a consistent, background-independent theory of quantum gravity.

It would be interesting to apply these concepts to higher-dimensional spacetime with constant curvature (e.g. AdS$_{2+1}$), looking forward to a more realistic description of scalar field propagation in non-trivial manifolds. Another field of interest would be the construction of a fermionic field propagator on a causal set. The problem is that there exists no analogous description of spinors on a causal set. However, as discussed in \cite{johnston2010quantumfieldscausalsets}, the goal is not necessarily to develop spinor models on a causal set, but rather to construct a model for spin-$\frac{1}{2}$ particles. This might require a complete reformulation of the standard continuum-based spinor theory, replacing it with an alternative, equally effective framework but capable to generalization on a causal set. One idea would be to investigate \textit{Feynman's Checker Model} \cite{feynman2010quantum}, where a particle moves at light speed on a discrete spacetime grid, zigzagging back and forth. By performing the sum over all possible paths with appropriate phase factors, the model reproduces the Dirac equation in $(1+1)$ dimensions. If and how exactly this can be implemented in a causal set setting is a current field of study.

\begin{acknowledgments}
The authors thank J. Erdmenger and R. Meyer for valuable discussions and suggestions.
\end{acknowledgments}

\appendix
\section{Supplementary Graphs}
\label{appendix}
%
\begin{figure}[H]
\centering
\subfigure[]{\includegraphics[width=0.45\textwidth]{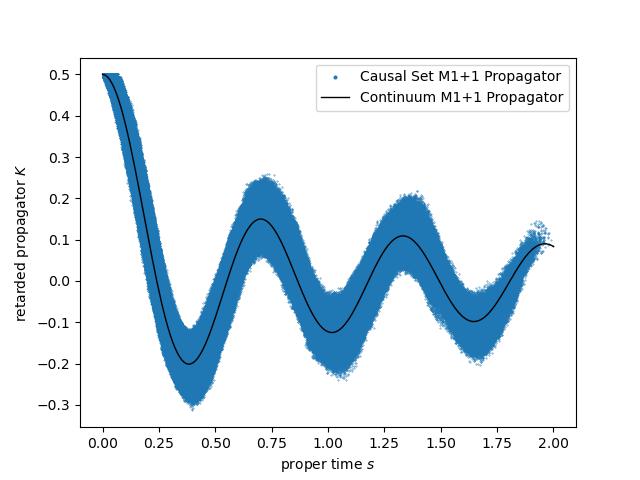}}
\hspace{0.0001\textwidth}
\subfigure[]{\includegraphics[width=0.45\textwidth]{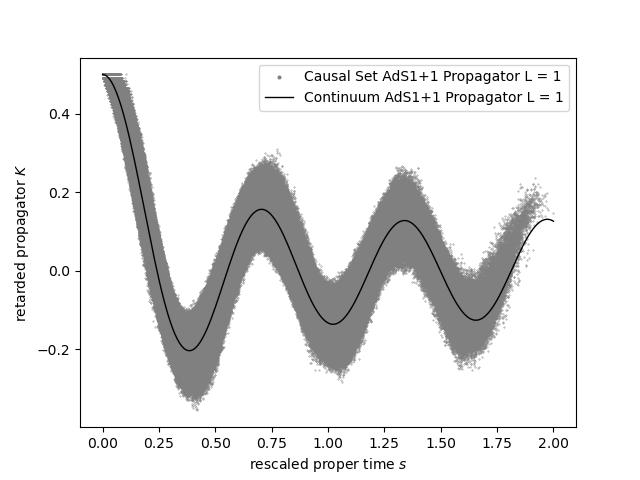}}
\caption{\justifying
Numerical results for the retarded causal set propagator (\ref{cstprop}) with $m=10$ for flat 2D Minkowski (a) and 2D anti de-Sitter spacetime with curvature radius $L=1$ from a single sprinkling with $n=18000$, together with the corresponding continuum solutions, all plotted over the (rescaled) manifold proper time.}
\label{fig:boringpropsfigs}
\end{figure}

\providecommand{\noopsort}[1]{}\providecommand{\singleletter}[1]{#1}%

\end{document}